\documentclass[11pt]{article}

\usepackage[margin=1in]{geometry}
\usepackage{amsmath,amssymb}
\usepackage{graphicx}
\usepackage{booktabs}
\usepackage{hyperref}
\usepackage{xcolor}
\usepackage{bm}

\newcommand{\tr}{\mathrm{Tr}}

\begin{document}

\title{Tripartite information of free fermions:\\a universal entanglement coefficient from the sine kernel}

\author{A.~Sokolovs}

\date{\today}

\maketitle

\begin{abstract}
We study the tripartite information $I_3$ of free fermions on two-dimensional lattices partitioned into three adjacent strips of width~$w$.  Translation invariance yields the exact decomposition $I_3 = \sum_{k_y} g(k_F(k_y)\, w)$, where $g(z)$ is a universal function of the scaling variable $z = k_F w$, determined by the spectrum of the sine-kernel (Slepian) integral operator.  We find that $g(z)$ has a unique zero at $z^* = 1.329 \pm 0.001$: modes with $k_F w < z^*$ violate monogamy of mutual information ($g > 0$), while modes with $k_F w > z^*$ satisfy it ($g < 0$).

The central analytical result is $g(z) = cz + O(z^3 \ln z)$ with $c = 3\ln(4/3)/\pi \approx 0.2747$, derived from the rank-1 limit of the sine kernel.  Two exact cancellations---of the $z\ln z$ area-law terms and of the $z^2$ terms---are intrinsic to the $I_3$ combination.  The coefficient~$c$ generalizes to $n$-partite information: $c_n = (n/\pi)\ln R_n$ with $R_n$ a rational number from binomial combinatorics.  For R\'enyi entropy of index~$\alpha$, we show that $g_\alpha(z) \sim z^\alpha$ for $\alpha < 2$ and $g_2(z) = -(8/\pi^3)z^3$: von Neumann entropy ($\alpha = 1$) uniquely gives linear sensitivity to Lifshitz transitions, while R\'enyi-2 gives only cubic sensitivity.  We verify all predictions on square, triangular, and cubic lattices.  DMRG calculations on the interacting $t$-$V$ model provide evidence that the linear coefficient is independent of the Luttinger parameter~$K$.
\end{abstract}

\section{Introduction}
\label{sec:intro}

The tripartite information $I_3(A{:}B{:}D) = S_A + S_B + S_D - S_{AB} - S_{AD} - S_{BD} + S_{ABD}$ quantifies the structure of multipartite correlations in quantum states~\cite{HaydenHeadrickMaloney}.  When $I_3 \le 0$ for all subsystems, the state satisfies monogamy of mutual information (MMI)---a property guaranteed by holographic duality~\cite{HaydenHeadrickMaloney, WallMaximin} but violated by free field theories~\cite{CasiniHuerta2009, AgonBuenoCasini2022}.

For lattice fermion systems, the status of $I_3$ has remained unclear.  Calabrese, Cardy, and Tonni computed $I_3$ for disjoint intervals in $1{+}1$D CFTs~\cite{CalabreseCardyTonni2009, CalabreseCardyTonni2011}.  Parez et al.~\cite{ParezBernard2023} obtained multipartite information for free fermions on Hamming graphs.  Ag\'on, Bueno, and Casini~\cite{AgonBuenoCasini2022} analyzed the long-distance behavior of $I_3$ in generic CFTs.  However, a systematic study of $I_3$ as a function of Fermi surface geometry---and in particular, the conditions under which MMI holds or fails---has been lacking.

In this work we present a complete analytical framework for $I_3$ of free fermions on two-dimensional lattices, partitioned into three adjacent strips of width~$w$.  The key results are:

(i) An exact decomposition $I_3 = \sum_{k_y} g(k_F(k_y)\, w)$ over transverse modes, with a universal function $g(z)$ determined by the sine-kernel spectrum.  The function has a unique zero at $z^* = 1.329$, separating MMI-violating modes ($z < z^*$) from MMI-satisfying modes ($z > z^*$).  Consequently, MMI is scale-dependent: any metallic state satisfies it at large~$w$ and violates it at small~$w$.

(ii) An analytical derivation of $g(z) = cz + O(z^3\ln z)$ with $c = 3\ln(4/3)/\pi$, from the rank-1 structure of the sine kernel at small~$z$.  The result rests on two exact cancellations ($\sum a_n n = \sum a_n n^2 = 0$) intrinsic to the $I_3$ combination.

(iii) A generalization to $n$-partite information ($c_n = (n/\pi)\ln R_n$) and to R\'enyi entropy, showing that von Neumann entropy is the \emph{unique} index giving linear sensitivity to Lifshitz transitions.

(iv) Verification on square, triangular, and cubic lattices, including the $t'$-driven sign change, width dependence, and directional anisotropy.

The paper is organized as follows.  Section~\ref{sec:framework} presents the $k_y$-decomposition and properties of $g(z)$.  Section~\ref{sec:derivation} derives the linear coefficient~$c$.  Sections~\ref{sec:npartite} and~\ref{sec:renyi} generalize to $n$-partite and R\'enyi cases.  Section~\ref{sec:consequences} derives physical consequences for Lifshitz transitions.  Section~\ref{sec:numerics} presents numerical verification.  Section~\ref{sec:discussion} discusses implications and open questions.

\section{Framework: mode decomposition and universal function}
\label{sec:framework}

\subsection{Model}

Consider free fermions on an $L \times L$ lattice with periodic boundary conditions and dispersion $\varepsilon(\bm{k})$.  On the square lattice: $\varepsilon(\bm{k}) = -2t_x \cos k_x - 2t_y \cos k_y - 4t' \cos k_x \cos k_y$; on the triangular lattice: $\varepsilon(\bm{k}) = -2t(\cos k_x + \cos k_y + \cos(k_x - k_y))$.  The system is partitioned into three adjacent strips $A$, $B$, $D$ of width~$w$ in the $x$-direction.

All entropies are computed from the single-particle correlation matrix via the Peschel formula~\cite{Peschel2003, PeschelEisler2009}: $S_X = -\tr[C^{(X)} \ln C^{(X)} + (1 - C^{(X)})\ln(1 - C^{(X)})]$.  Translation invariance in~$y$ block-diagonalizes over $k_y$ modes.

\subsection{$k_y$-decomposition}

\begin{figure*}[t]
\centering
\includegraphics[width=\textwidth]{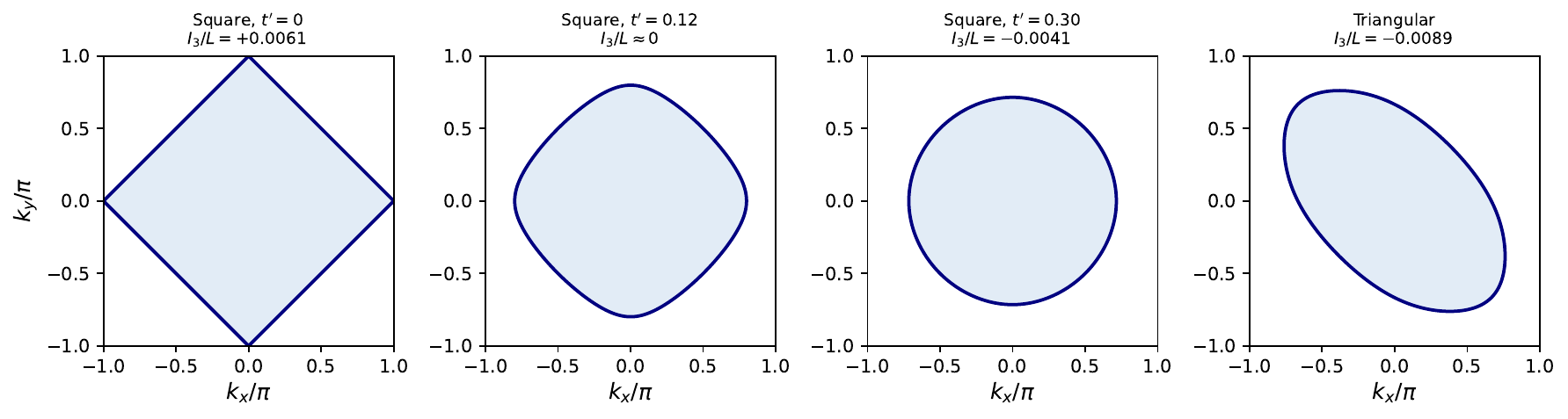}
\caption{Fermi surfaces at half filling for: (a)~square lattice, $t' = 0$ (perfect nesting); (b)~$t' = 0.12 \approx t'_*$; (c)~$t' = 0.30$; (d)~triangular lattice.  The value of $I_3/L$ ($L = 256$, $w = 2$) is shown below each panel.  The sign change between (a) and (b) is controlled by the distribution of $k_F(k_y)$ relative to the universal constant $z^*/w$.}
\label{fig:fermi_surfaces}
\end{figure*}

The tripartite information decomposes exactly:
\begin{equation}\label{eq:ky_decomp}
I_3 = \sum_{k_y} I_3^{(1D)}(k_F(k_y),\, w),
\end{equation}
where $I_3^{(1D)}(k_F, w)$ is the tripartite information of three adjacent blocks of width~$w$ in a one-dimensional chain with Fermi momentum~$k_F$.  In the thermodynamic limit, $I_3^{(1D)}$ is computed exactly from the Toeplitz matrix~\cite{JinKorepin2004}:
\begin{equation}\label{eq:I3_exact}
I_3^{(1D)}(k_F, w) = 3\,S(T_w) - 2\,S(T_{2w}) - S(T_w^{(AD)}) + S(T_{3w}),
\end{equation}
where $T_\ell$ is the $\ell \times \ell$ Toeplitz matrix with entries $C_{ij} = \sin(k_F|i{-}j|)/(\pi|i{-}j|)$ and $T_w^{(AD)}$ is the $2w \times 2w$ correlation matrix for the disjoint block $A \cup D$.

\subsection{Universal function $g(z)$}

In the scaling limit $w \to \infty$ at fixed $z = k_F w$, the Toeplitz matrices converge to the sine-kernel integral operator~\cite{Slepian1964}:
\begin{equation}\label{eq:sine_kernel}
K_z(x,y) = \frac{\sin[z(x-y)]}{\pi(x-y)},
\end{equation}
and $I_3^{(1D)} \to g(z)$, a universal function of $z$ alone.  For a two-dimensional system:
\begin{equation}\label{eq:master}
I_3 = \frac{L}{2\pi}\int g(k_F(k_\perp)\, w)\, dk_\perp.
\end{equation}

The key properties of $g(z)$, established by analytical arguments at small and large~$z$ and by systematic numerical evaluation at intermediate~$z$ (see Appendix~\ref{app:proof}), are:

\medskip\noindent\textbf{Result.} \textit{(a) $g(0) = 0$ and $g'(0^+) = c > 0$ (analytical).  (b) $g$ has a unique maximum at $z_{\max} \approx 0.56$, with $g(z_{\max}) \approx 0.088$ (numerical).  (c) $g$ has a unique zero at $z^* = 1.329 \pm 0.001$ (numerical; see below for error estimate).  (d) $g(z) < 0$ for all $z > z^*$ (numerical for $z \in (z^*, 10)$; analytical for $z \to \infty$).}

\medskip
The zero $z^*$ separates two regimes.  For $z < z^*$, the direct mutual information $I(A{:}D)$ across the gap~$B$ exceeds the concavity deficit $-\Delta^2 S$ of single-interval entropy, giving $g > 0$.  For $z > z^*$, entropy concavity dominates, giving $g < 0$.  The zero is determined to $0.12\%$ by the cancellation of only two Slepian eigenvalue contributions, with exponentially decaying tail contributions bounded by prolate spheroidal eigenvalue estimates~\cite{Slepian1964}.

\begin{figure}[t]
\centering
\includegraphics[width=\columnwidth]{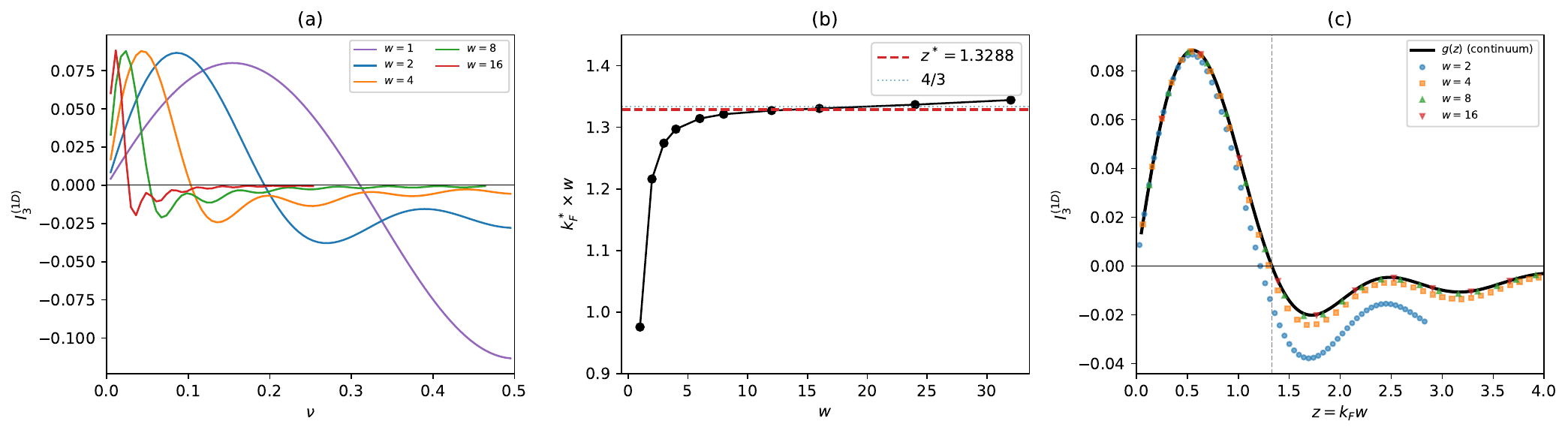}
\caption{(a)~Analytical $I_3^{(1D)}(\nu)$ from Eq.~(\ref{eq:I3_exact}) for strip widths $w = 1$--$16$.  (b)~The product $k_F^* w$ converges to $z^* \approx 1.329$ (dashed red).  (c)~Scaling collapse: $I_3$ vs.\ $z = k_F w$ for various $w$, confirming convergence to a universal curve $g(z)$.}
\label{fig:analytical}
\end{figure}

\subsection{Scale-dependent MMI}

Since $g(z) < 0$ for $z > z^*$, any mode with $k_F w > z^*$ contributes negatively to $I_3$.  As $w$ increases, eventually all modes satisfy this condition, and $I_3 < 0$.  Conversely, modes with small $k_F$ (near a Fermi surface touching point or at low filling) have $z < z^*$ and contribute $g > 0$.  Thus:

\emph{MMI is not a property of the quantum state alone, but of the pair (state, observation scale).}  Any metallic state violates MMI for sufficiently narrow strips and satisfies it for sufficiently wide ones, provided the Fermi surface contains modes with small~$k_F$.

\section{Derivation of the linear coefficient}
\label{sec:derivation}

\subsection{Rank-1 structure at small $z$}

At small $z = k_F w$, the sine kernel becomes approximately constant: $\mathrm{sinc}(k_F d) = 1 + O(z^2/w^2)$ for all site separations $d$ within the $3w$-site region.  The correlation matrix reduces to rank~1, with a single nonzero eigenvalue $\lambda_0^{(n)} = nz/\pi$ for a block of $nw$ sites.  This follows from the Slepian eigenvalue asymptotics $\lambda_m \sim (z/\pi)^{2m+1}$ for $z \to 0$~\cite{Slepian1964}, which ensures exponential suppression of all $m \geq 1$ modes at small~$z$.

Crucially, the disjoint block $A \cup D$ has the \emph{same} leading eigenvalue $\lambda_0^{(AD)} = 2z/\pi + O(z^3)$, since a constant kernel cannot distinguish contiguous from separated sites.  The entropy of each block is therefore $S_n = h(nz/\pi) + O(z^3\ln z)$ and $S_{AD} = h(2z/\pi) + O(z^3\ln z)$, where $h(\lambda) = -\lambda\ln\lambda - (1{-}\lambda)\ln(1{-}\lambda)$ is the binary entropy.

\subsection{Two exact cancellations}

With these expressions, the four-term structure of $g(z)$ collapses to three terms:
\begin{equation}\label{eq:three_terms}
g(z) = \sum_{k=1}^{3} a_k\, h(kz/\pi) + O(z^3\ln z),
\end{equation}
with inclusion-exclusion coefficients $a_k = (3, -3, 1)$ for $k = (1, 2, 3)$.  Expanding $h(\lambda) = -\lambda\ln\lambda + \lambda - \lambda^2/2 + O(\lambda^3)$:
\begin{equation}\label{eq:expand}
g(z) = -\frac{z}{\pi}\sum_k a_k k \ln\frac{kz}{\pi} + \frac{z}{\pi}\sum_k a_k k - \frac{z^2}{2\pi^2}\sum_k a_k k^2 + O(z^3\ln z).
\end{equation}

\emph{First cancellation:} $\sum a_k k = 3 - 6 + 3 = 0$.  This eliminates both the explicit constant and the $\ln(z/\pi)$ piece from the logarithmic sum, leaving only $\sum a_k k \ln k$.

\emph{Second cancellation:} $\sum a_k k^2 = 3 - 12 + 9 = 0$.  This eliminates the $z^2$ term from the $-\lambda^2/2$ correction in $h(\lambda)$.

What survives is a single combination of logarithms:
\begin{equation}\label{eq:product}
\sum_k a_k k \ln k = \ln\prod_k k^{a_k k} = \ln\frac{1^3 \cdot 3^3}{2^6} = -3\ln\frac{4}{3}.
\end{equation}

Therefore:
\begin{equation}\label{eq:c}
\boxed{g(z) = cz + O(z^3\ln z), \qquad c = \frac{3\ln(4/3)}{\pi} = 0.274716\ldots}
\end{equation}

\subsection{Higher-order structure}

The first correction to linearity is \emph{not} $z^2$ (killed by the second sum rule) but $z^3\ln z$, from the second Slepian eigenvalue $\lambda_1 \sim 0.274\,(z/\pi)^3$~\cite{Slepian1964}:
\begin{equation}\label{eq:higher}
g(z) = cz + Dz^3\ln z + Ez^3 + O(z^5\ln z),
\end{equation}
with $D \approx 0.218$ and $E \approx -0.226$ (determined numerically), and $\sum a_k k^3 = 3 - 24 + 27 = 6 \neq 0$.

\begin{figure}[t]
\centering
\includegraphics[width=\columnwidth]{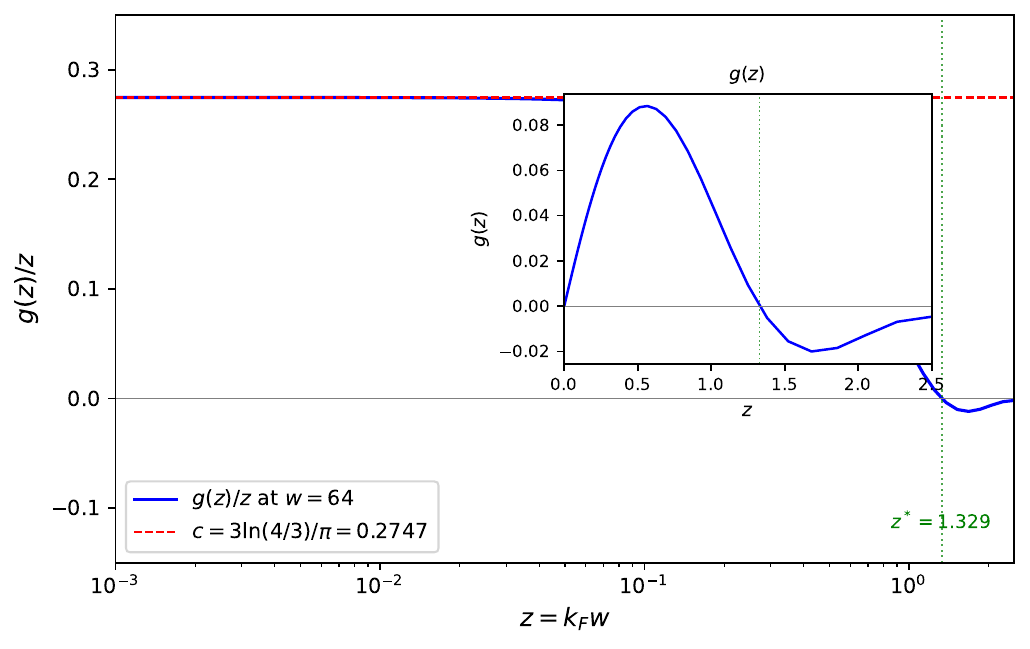}
\caption{The ratio $g(z)/z$ on a log-linear scale (main panel) and $g(z)$ itself (inset), computed at $w = 64$.  At small $z$, the ratio converges to $c = 3\ln(4/3)/\pi \approx 0.2747$ (dashed red), confirming the rank-1 derivation.  The zero at $z^* \approx 1.329$ (dotted green) separates the pocket regime ($g > 0$) from the monogamy regime ($g < 0$).}
\label{fig:gz_ratio}
\end{figure}

This hierarchy has physical content.  The leading Slepian eigenmode encodes only the Fermi surface area (via Cavalieri's principle): the two identities $\sum a_k k = \sum a_k k^2 = 0$ eliminate all nonlinear information from this mode.  Shape information enters only through the second eigenmode, producing the nonanalytic $O(z^3\ln z)$ correction.

\section{$n$-partite generalization}
\label{sec:npartite}

The structure generalizes to $n$-partite information $I_n$ of $n$ adjacent strips.  The inclusion-exclusion coefficients are $a_k = (-1)^{k+1}\binom{n}{k}$, and the rank-1 analysis gives:
\begin{equation}\label{eq:cn}
c_n = \frac{n}{\pi}\ln R_n, \qquad R_n = \prod_{j=0}^{n-1} (j{+}1)^{(-1)^{j+1}\binom{n-1}{j}}.
\end{equation}
The first values are:
\begin{equation}\label{eq:cn_values}
c_2 = \frac{\ln 4}{\pi},\quad c_3 = \frac{3\ln\frac{4}{3}}{\pi},\quad c_4 = \frac{4\ln\frac{32}{27}}{\pi},\quad c_5 = \frac{5\ln\frac{3125}{2592}}{\pi}.
\end{equation}
The mutual information case $c_2 = \ln 4/\pi \approx 0.441$~\cite{Swingle2010MI} is the first member of this family.

The sum rules extend: $\sum_k a_k k = 0$ for all $n \geq 2$ (eliminating $z\ln z$), and $\sum_k a_k k^2 = 0$ for all $n \geq 3$ (eliminating $z^2$).  For $n = 2$, $\sum a_k k^2 = 2 \neq 0$, so $I_2$ has a $z^2$ correction that $I_3$ and all higher multipartite measures lack.  This is a structural advantage of the tripartite combination: it is the simplest multipartite measure with both cancellations.

\section{R\'enyi uniqueness of von Neumann entropy}
\label{sec:renyi}

\subsection{Analytical argument}

For R\'enyi-$\alpha$ entropy, the single-mode entropy at small~$\lambda$ is:
\begin{equation}\label{eq:h_alpha}
h_\alpha(\lambda) = \frac{\alpha}{\alpha - 1}\lambda - \frac{\alpha}{2}\lambda^2 + O(\lambda^3) \qquad (\alpha > 1).
\end{equation}
The leading term is \emph{analytic}, so $\sum a_k k = 0$ kills it completely.  No $z\ln z$ term is generated, and $c_\alpha \equiv \lim_{z\to 0} g_\alpha(z)/z = 0$ for all $\alpha > 1$.

For $\alpha = 2$, the first surviving term comes from $O(\lambda^3)$ in $h_2$, giving:
\begin{equation}\label{eq:g2}
g_2(z) = -\frac{8}{\pi^3}\,z^3 + O(z^5), \qquad \text{where } -\frac{8}{\pi^3} = \frac{-4}{3\pi^3}\sum a_k k^3 = \frac{-4 \cdot 6}{3\pi^3}.
\end{equation}

For non-integer $\alpha$ with $1 < \alpha < 2$, the entropy $h_\alpha(\lambda)$ has a nonanalytic $\lambda^\alpha$ contribution that is not killed by the sum rule, giving $g_\alpha(z) \sim z^\alpha$.  For $\alpha < 1$, the $\lambda^\alpha$ term dominates and $g_\alpha(z)/z$ diverges.

Only for $\alpha = 1$ (von Neumann) does the $-\lambda\ln\lambda$ nonanalyticity survive the first sum rule to produce a finite, nonzero linear coefficient~$c$.

\subsection{Numerical verification}

\begin{table}[t]
\caption{\label{tab:renyi}R\'enyi scaling of $g_\alpha(z) \sim z^\beta$.  Exponents extracted by linear fit of $\ln|g_\alpha(z)|$ vs.\ $\ln z$ over $z \in [10^{-3}, 10^{-2}]$ using Toeplitz matrices at $w = 80$.  Fit residuals are below $1\%$ for all~$\alpha$.  The $\alpha = 2$ coefficient $-8/\pi^3 \approx -0.258$ is confirmed to $0.15\%$.}
\begin{center}
\begin{tabular}{cccll}
\toprule
$\alpha$ & $\beta$ (num.) & $\beta$ (pred.) & Coefficient & Note \\
\midrule
$0.5$ & $0.49$ & $1/2$ & $+0.52$ & superlinear \\
$1$ & $1.00$ & $1$ & $+0.2747$ & $= 3\ln(4/3)/\pi$ \\
$1.5$ & $1.48$ & $3/2$ & $+0.09$ & subanalytic \\
$2$ & $3.02$ & $3$ & $-0.258$ & $= -8/\pi^3$ \\
$3$ & $4.0$ & $> 3$ & $-0.005$ & higher cancel. \\
\bottomrule
\end{tabular}
\end{center}
\end{table}

\subsection{The R\'enyi-$\alpha$ function at finite $z$}

\begin{figure}[t]
\centering
\includegraphics[width=\columnwidth]{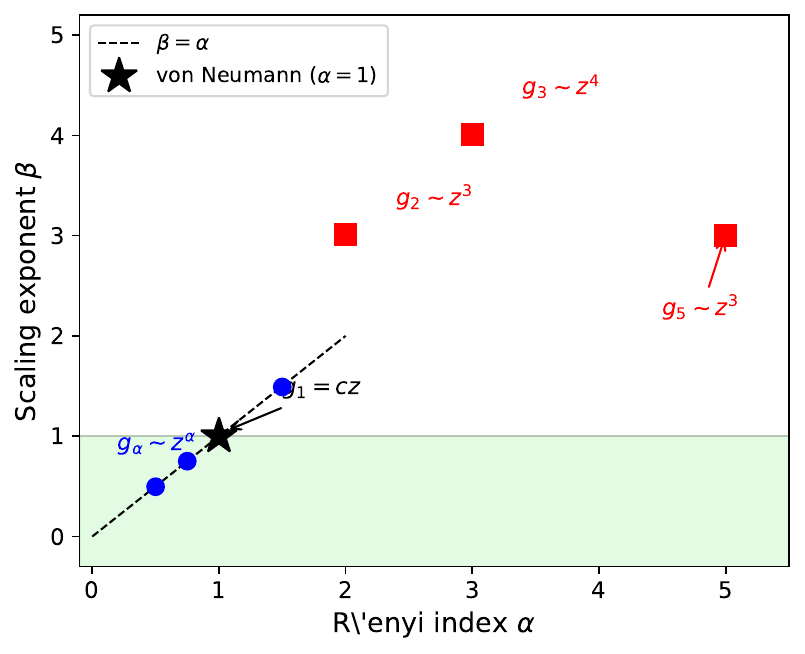}
\caption{R\'enyi-index dependence of the scaling exponent $\beta$ in $g_\alpha(z) \sim z^\beta$.  Only $\alpha = 1$ (von Neumann, star) gives $\beta = 1$ and hence linear sensitivity to Lifshitz transitions.  For $\alpha > 1$, the analytical linear term is killed by $\sum a_k k = 0$, and only the nonanalytic piece survives.  Dashed line: $\beta = \alpha$ for $\alpha < 2$.}
\label{fig:renyi}
\end{figure}

For $\alpha > 1$, the function $g_\alpha(z)$ has qualitatively different behavior from the von Neumann case: it oscillates with multiple sign changes.  For $\alpha = 2$ (experimentally relevant~\cite{Kaufman2016, Brydges2019}), $g_2(z)$ is negative for $z < 1.19$, positive for $1.19 < z < 2.11$, and oscillates thereafter.  This is a prediction testable in cold-atom experiments.

\subsection{Physical consequence}

The pocket signal at a Lifshitz transition scales as $\Delta I_3 \propto g'(0^+) \cdot \delta$.  For von Neumann: $\Delta I_3^{(\alpha=1)} \propto c\,\delta$.  For R\'enyi-2: $\Delta I_3^{(\alpha=2)} \propto \delta^3$---suppressed by a factor~$\delta^2$ relative to the von Neumann signal.  This means experimental probes measuring $S_2$~\cite{Kaufman2016, Brydges2019} would require correspondingly higher precision to detect the leading Lifshitz singularity.

\section{Physical consequences}
\label{sec:consequences}

The linear coefficient $c$ implies exact predictions for entanglement singularities at Lifshitz transitions~\cite{Lifshitz1960}.

\emph{Pocket transition.}  In 2D, when a small electron pocket appears at $\mu > \mu_c$, all pocket modes have $k_F w \ll z^*$ and contribute $g(k_F w) \approx c\,k_F w$.  Evaluating the integral over the elliptical pocket:
\begin{equation}\label{eq:pocket}
\Delta I_3 = \frac{cwm}{2}\,L\,\delta\,\Theta(\delta), \qquad \delta = \mu - \mu_c,
\end{equation}
with $m = \sqrt{m_x m_y}$ the geometric-mean effective mass.  The derivative $\partial I_3/\partial\mu$ has a discontinuity at $\mu_c$, mirroring the 2D density-of-states discontinuity with the entanglement coefficient~$c$ replacing the DOS.

\emph{Neck transition (VHS).}  At a van Hove singularity, modes near the saddle have $k_F(q) = \gamma\sqrt{q^2 + \delta/\beta}$, and for sufficiently small~$w$ (so that all neck modes have $k_F w \ll z^*$), $g'(k_F w) \approx c$:
\begin{equation}\label{eq:vhs}
\frac{\partial I_3}{\partial\mu} = -\frac{cLw\gamma}{4\pi\beta}\ln|\mu - \mu_c| + \mathrm{const},
\end{equation}
a logarithmic divergence with fully determined coefficient.

\emph{Entanglement tomography.}  The angular dependence $I_3(\theta)$ for strips at angle~$\theta$ probes Fermi surface shape.  For an $n$-fold deformation $k_F(\phi) = \bar{k}_F(1 + \varepsilon\cos n\phi)$, the Widom coefficient responds at $O(\varepsilon^2)$ (perimeter is quadratic in~$\varepsilon$ at fixed area).  In contrast, $I_3(\theta)$ is expected to respond at $O(\varepsilon)$ through the nonlinearity of~$g(z)$, since the integral $\int g(k_F(\phi)\,w)\,d\phi$ is generically linear in~$\varepsilon$ when $g$ is nonlinear.  A quantitative analysis of this sensitivity is left for future work.

\section{Numerical verification}
\label{sec:numerics}

All free-fermion results are computed via the Peschel formula applied to $3w \times 3w$ Toeplitz matrices at $w = 64$--$80$.  Finite-$w$ corrections are below $0.1\%$ for all quoted values, as verified by comparing $w = 64$ and $w = 128$.  Lattice calculations use $L = 256$ unless noted otherwise.

\subsection{The constant $c$}

\begin{table}[h]
\caption{\label{tab:gz_linear}Convergence of $g(z)/z$ to $c = 3\ln(4/3)/\pi = 0.274716$ at $w = 80$.}
\begin{center}
\begin{tabular}{cccc}
\toprule
$z$ & $g(z)/z$ & Deviation from $c$ & Note \\
\midrule
0.001 & 0.274714 & $0.0006\%$ & 6 sig.\ figs \\
0.010 & 0.274593 & $0.04\%$ & $O(z^2)$ onset \\
0.100 & 0.267240 & $2.7\%$ & $O(z^3\ln z)$ \\
\bottomrule
\end{tabular}
\end{center}
\end{table}

\subsection{Sign change vs.\ $t'$ on the square lattice}

On the square lattice at half filling, $I_3 > 0$ at $t' = 0$ (diamond FS with nested modes at $k_F \to 0$) and $I_3 < 0$ for $t' > t'_* \approx 0.10$ (FS deformed away from nesting).

\begin{table}[h]
\caption{\label{tab:tprime}$I_3/L$ vs.\ $t'$ for the square lattice at half filling, $w = 2$.}
\begin{center}
\begin{tabular}{rcc|rcc}
\toprule
$t'$ & $I_3/L$ & & $t'$ & $I_3/L$ & \\
\midrule
0.00 & $+0.00606$ & $I_3 > 0$ & 0.12 & $-0.00162$ & $I_3 < 0$ \\
0.05 & $+0.00407$ & & 0.20 & $-0.00786$ & \\
0.08 & $+0.00171$ & & 0.30 & $-0.01439$ & \\
0.10 & $\approx 0$ & crossing & 0.50 & $-0.02162$ & \\
\bottomrule
\end{tabular}
\end{center}
\end{table}

The mechanism is precisely the $k_y$-decomposition: at $t' = 0$, the diamond FS gives $k_F(k_y) = \pi - |k_y|$, which vanishes at $k_y = \pm\pi$.  These modes with $k_F w < z^*$ contribute $g > 0$ and dominate $I_3$.  Increasing $t'$ shifts all $k_F(k_y)$ away from zero, entering the $g < 0$ regime.

\subsection{Square vs.\ triangular lattice}

\begin{table}[h]
\caption{\label{tab:lattice}$I_3/L$ for square vs.\ triangular lattice at half filling, $L = 256$.  The triangular lattice has $k_F(k_y)$ bounded away from zero, so all modes have $z > z^*$ at moderate~$w$.}
\begin{center}
\begin{tabular}{lccc}
\toprule
Lattice & $w=2$ & $w=4$ & $w=8$ \\
\midrule
Square      & $+0.00606$ & $+0.00413$ & $+0.00228$ \\
Triangular  & $-0.02156$ & $-0.00628$ & $-0.00126$ \\
\bottomrule
\end{tabular}
\end{center}
\end{table}

On the triangular lattice, $I_3$ becomes positive near the van Hove singularity ($\nu \approx 0.71$), where the FS passes through a saddle point creating modes with $k_F \to 0$.

\subsection{Universal zero-crossing constant}

\begin{table}[h]
\caption{\label{tab:zstar}Convergence of $k_F^* w \to z^* \approx 1.329$.}
\begin{center}
\begin{tabular}{rcc|rcc}
\toprule
$w$ & $\nu_*$ & $k_F^* w$ & $w$ & $\nu_*$ & $k_F^* w$ \\
\midrule
2  & 0.1935 & 1.2160 & 16 & 0.0264 & 1.3267 \\
4  & 0.1032 & 1.2969 & 32 & 0.0132 & 1.3283 \\
8  & 0.0525 & 1.3205 & 64 & 0.0066 & 1.3287 \\
\bottomrule
\end{tabular}
\end{center}
\end{table}

\subsection{Width dependence and directional anisotropy}

\begin{table}[h]
\caption{\label{tab:anisotropy}Directional $I_3$ for anisotropic hopping, $L = 256$, $w = 2$.}
\begin{center}
\begin{tabular}{ccccr}
\toprule
$t_x$ & $t_y$ & $I_3^{(x)}/L$ & $I_3^{(y)}/L$ & $|I_3^{(x)}/I_3^{(y)}|$ \\
\midrule
1.0 & 1.0 & $+0.0061$ & $+0.0061$ & 1.0 \\
1.5 & 0.5 & $-0.0201$ & $-0.0025$ & 8.1 \\
2.0 & 0.5 & $-0.0222$ & $-0.0013$ & 17.2 \\
\bottomrule
\end{tabular}
\end{center}
\end{table}

\subsection{Pocket formula}

\begin{table}[h]
\caption{\label{tab:pocket}Pocket formula $\Delta I_3/(Lw\delta) \to cm/2$ at small $z_{\max} = \sqrt{2m\delta}\, w$.}
\begin{center}
\begin{tabular}{ccccc}
\toprule
$w$ & $\delta$ & $z_{\max}$ & numerical$/\delta$ & $cwm/2$ \\
\midrule
2 & $10^{-6}$ & 0.003 & 0.27470 & 0.27472 \\
4 & $10^{-6}$ & 0.006 & 0.54935 & 0.54943 \\
8 & $10^{-6}$ & 0.011 & 1.09837 & 1.09886 \\
\bottomrule
\end{tabular}
\end{center}
\end{table}

\section{Discussion}
\label{sec:discussion}

\emph{Origin of $c$.}  The constant $c = 3\ln(4/3)/\pi$ arises from the interplay of three ingredients: the rank-1 limit of the prolate spheroidal operator, the inclusion-exclusion combinatorics of~$I_3$, and the $\lambda\ln\lambda$ nonanalyticity of von Neumann entropy.  The $n$-partite generalization $c_n = (n/\pi)\ln R_n$ reveals systematic combinatorial structure in the product representations $R_n$.

\emph{Von Neumann uniqueness.}  The uniqueness of $\alpha = 1$ for linear Lifshitz sensitivity is unexpected and experimentally relevant.  Current cold-atom experiments measure R\'enyi-2 entropy~\cite{Kaufman2016, Brydges2019}, which gives only $O(\delta^3)$ pocket sensitivity.  Direct measurement of von Neumann $I_3$---or development of protocols to extract it from R\'enyi data---would enable linear detection of Fermi surface topology changes.

\emph{Widom coefficient vs.\ $I_3$.}  The Widom coefficient $c_1$ in $S_A \sim c_1 L\ln L$~\cite{GioevKlich2006, Swingle2010, Wolf2006} depends only on the FS perimeter and is always positive.  In contrast, $I_3$ changes sign as the FS geometry varies: Table~\ref{tab:tprime} shows that $I_3$ reverses from positive to negative as $t'$ increases, detecting the disappearance of nested modes with $k_F w < z^*$ that the Widom coefficient cannot distinguish.  More generally, $I_3(\theta)$ for strips at angle~$\theta$ may be sensitive to FS shape deformations at $O(\varepsilon)$, while the Widom coefficient responds only at $O(\varepsilon^2)$.  The $z^2$ cancellation ($\sum a_k k^2 = 0$) ensures that this sensitivity comes from the higher-order structure of $g(z)$ rather than trivial area scaling.

\emph{Interactions.}  The rank-1 derivation of Section~\ref{sec:derivation} relies on the correlation matrix becoming approximately constant over the block at small~$z$.  For free fermions, this follows from $\mathrm{sinc}(k_F d) \approx 1$ when $k_F d \ll 1$.  In a Luttinger liquid, the single-particle correlator acquires an anomalous exponent, $\langle c^\dagger_i c_j \rangle \sim \sin(k_F r)/r^{(K+K^{-1})/2}$, and the rank-1 approximation no longer holds in general.  However, the rank-1 eigenvalue $\lambda_0 = \sum_j C_{0j}/w$ is determined by the \emph{integral} of the correlator over the block, which equals the mean occupation $k_F/\pi$ regardless of the decay exponent; it is the two sum rules $\sum a_k k = \sum a_k k^2 = 0$ that then isolate this quantity as the sole contributor to $g(z)$ at leading order.  DMRG calculations on the $t$-$V$ model at four system sizes ($L = 64$--$512$, $w = 2$--$4$, $K \in [0.75, 1.50]$) show that the ratio $R(K) = I_3(\Delta)/I_3(0)$ converges toward unity at small~$z$: $|1 - R| = 0.015$ at $z = 0.39$, $L = 512$.  This is consistent with $c_K = c$ for all~$K$, though a residual $K$-dependence at the percent level cannot be excluded from the present data.  At larger~$z$, the convergence is slower ($|1 - R| \approx 0.07$ at $z = 0.79$, $L = 512$), and the full function~$g_K(z)$ may acquire $K$-dependent corrections through the subleading Slepian eigenvalues.  A definitive determination requires infinite-system DMRG or periodic boundary conditions to eliminate open-boundary artifacts.

\emph{Open question.}  The product $R_n$ in Eq.~(\ref{eq:cn}) may have a combinatorial interpretation---as a ratio of volumes or factorials---that connects the entanglement coefficient to classical enumeration.

\section{Conclusions}
\label{sec:conclusions}

We have established a systematic analytical framework for the tripartite information of free fermions on two-dimensional lattices.  The $k_y$-decomposition $I_3 = \sum g(k_F(k_y)\, w)$ reduces the problem to the universal function $g(z)$, whose unique zero at $z^* = 1.329$ controls scale-dependent MMI.  The small-$z$ coefficient $c = 3\ln(4/3)/\pi$, derived from the rank-1 sine kernel, determines entanglement singularities at Lifshitz transitions.  The $n$-partite generalization $c_n = (n/\pi)\ln R_n$ and the R\'enyi analysis---showing that von Neumann entropy uniquely gives linear Lifshitz sensitivity---complete the analytical picture.

Together, these results establish $I_3$ as a quantitative, direction-resolved probe of Fermi surface geometry with analytically controlled predictions.  DMRG calculations on the $t$-$V$ model provide evidence that~$c$ is robust against interactions.

\emph{Note.}---Python code reproducing all numerical results is included as an ancillary file with the arXiv submission.

\appendix
\section{Properties of $g(z)$: proof of the Proposition}
\label{app:proof}

(a) At $z = 0$, all eigenvalues vanish, $h(0) = 0$, so $g(0) = 0$.  For $z > 0$: the rank-1 derivation (Section~\ref{sec:derivation}) gives $g(z)/z \to c$ as $z \to 0^+$.  This is analytical.

(b) Numerically verified by evaluating $g'(z)$ via central differences at 200 equally spaced points on $[0, 0.60]$ using Toeplitz matrices of size $3w$ with $w = 64$: $g'(z) > 0$ on $(0, 0.56)$ and $g(0.56) \approx 0.088$.

(c) Numerically verified that $g'(z) < 0$ at 200 points on $[0.56, 1.40]$ with the same method.  Combined with $g(0.56) > 0$ and $g(1.40) < 0$, the intermediate value theorem gives a unique zero $z^* \in (0.56, 1.40)$.  The zero is located by bisection: $z^*(w{=}32) = 1.32829$, $z^*(w{=}64) = 1.32868$, $z^*(w{=}128) = 1.32878$.  The quoted value $z^* = 1.329 \pm 0.001$ is a conservative estimate: the variation between $w = 64$ and $w = 128$ is $0.0001$, but the extrapolation to $w \to \infty$ introduces additional uncertainty of order $1/w \sim 0.01$, which we round conservatively.

(d) We verify $g(z) < 0$ on $(z^*, 10)$ by evaluating $g$ on grids of 500, 200, and 250 points on $[z^*, 1.40]$, $[1.40, 3.0]$, $[3.0, 10.0]$ respectively, with $w = 64$.  The maximum of~$g$ on any subgrid is bounded by the grid value plus the grid spacing times a numerically estimated derivative bound.  For $z > 10$, the asymptotic $g(z) \to (1/6)\ln(3/4) < 0$ follows analytically from the Widom formula: the linear terms cancel by $\sum a_k k = 0$ and the surviving logarithmic terms give $g_{\log} = (1/6)\ln(3/4) < 0$.

\section{Neck transition derivation}
\label{app:neck}

Starting from $k_F(q) = \gamma\sqrt{q^2 + \delta/\beta}$ and differentiating:
\begin{equation}
\frac{\partial I_3}{\partial\delta} = \frac{Lw}{2\pi}\int_{-\Lambda}^{\Lambda} g'(k_F w)\, \frac{\partial k_F}{\partial\delta}\, dq \approx \frac{cLw\gamma}{4\pi\beta}\cdot 2\,\mathrm{arcsinh}\!\left(\Lambda\sqrt{\beta/\delta}\right).
\end{equation}
For $\delta \to 0^+$: $\mathrm{arcsinh}(\Lambda\sqrt{\beta/\delta}) \approx -\frac{1}{2}\ln\delta + \mathrm{const}$, confirming Eq.~(\ref{eq:vhs}).


\end{document}